\documentclass[11pt]{article}

\usepackage{amsmath,amssymb}
\usepackage{graphicx}

\begin{document}

\begin{center}
{\LARGE Exploring Fifth Force Interactions with 18th Century Technology}\\
\vskip0.5in
{\Large Jason H. Steffen}\\
University of Washington at Seattle,\\ 
P.O.Box 351560,\\
Seattle, WA 98195-1560, USA\\
E-mail: jsteffen@u.washington.edu\\
\end{center}

\vskip0.3in

\begin{center}
\textsc{Abstract}
\end{center}
\textit{Many theories which unify gravity with the other known forces of nature predict the existence of an intermediate-range ``fifth force'' similar to gravity.  Such a force could be manifest as a deviation from the gravitational inverse-square law.  Currently, at distances near $10^{-1}$m, the inverse-square law is known to be correct to about one part per thousand.  I present the design of an experiment that will improve this limit by two orders of magnitude.  This is accomplished by constructing a torsion pendulum and source mass apparatus that are particularly insensitive to Newtonian gravity and, simultaneously, maximally sensitive to violations of the same.}

%\vskip3.5in
%\scriptsize{This essay received an "honorable mention" in the 2004 Essay Competition of the Gravity Research Foundation -- Ed.}

\newpage

\begin{figure}[!ht]
\begin{center}
\includegraphics[width=0.6\textwidth]{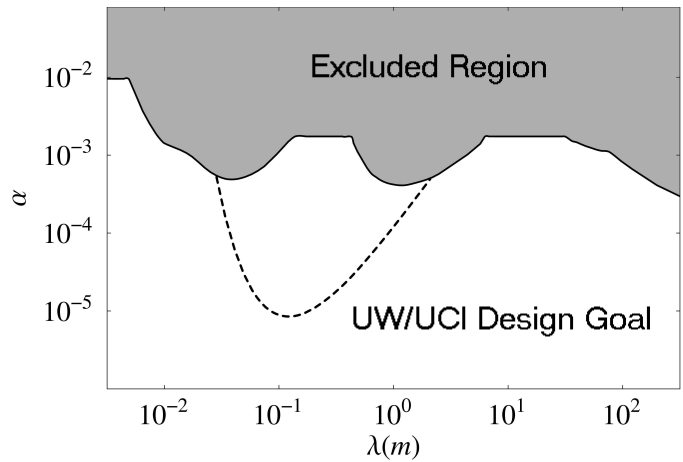}
\caption{Composition-independent experimental limit on $\alpha$ as a function of $\lambda$.  The dashed line denotes the design goal of the experiment described in this paper.  The experiment is a collaboration between the physics departments at the University of Washington and the University of California, Irvine.}
\label{alpha}
\end{center}
\end{figure}

Efforts by theoreticians to unify gravity with the other known forces of nature, to study physics at the Planck scale, or to extend the standard model predict various massive scalar or vector bosonic fields\cite{paul,ephr}.  Such fields could couple to conserved quantum numbers such as baryon number or to mass-energy and would, therefore, mimic a gravitational interaction.  A common parameterization of the effects of such interactions adds a Yukawa potential to the weak-field gravitational potential giving
\begin{equation}\label{potential}
\begin{split}
V &= V_{\text{grav}}+V_{\text{Yukawa}}\\
&=-\frac{Gm}{r}\left( 1 + \alpha e^{-\lambda/r} \right)
\end{split}
\end{equation}
where $\alpha$ is the strength of the additional potential relative to gravity and $\lambda$ is a characteristic length scale which depends upon the mass of the mediating boson.  The presence of this additional potential violates the inverse-square law of gravity (ISLV).  Its detection, or lack thereof, places constraints on the allowed theories.  At length scales between $\lambda = 10^{-3}$m and $\lambda = 10^{2}$m the current upper limit of the strength of this additional potential is near $\alpha \lesssim 10^{-3}$ (see Figure \ref{alpha}).

Improvements upon these limits require an experimental device capable of detecting extremely small forces.  For this reason, the torsion pendulum is often the instrument of choice.  From its first use in gravitation during the 18th century until now the basic design remains the same.  A pendulum mass, suspended by a thin fiber, is placed in the presence of a source mass.  Interactions between the pendulum and source cause a detectable change in the behavior of the pendulum such as a change in the pendulum equilibrium position, oscillation frequency, or amplitude of the second harmonic\cite{paul}.

I present the design of a null experiment that is expected to constrain $\alpha \lesssim 10^{-5}$ at a length scale of $\lambda = 10^{-1}$m, a two order-of-magnitude improvement over the current limit at that scale.  The experiment uses a torsion pendulum specifically designed to be insensitive to standard gravity and, at the same time, maximally sensitive to ISLV.  The novelty of designing a pendulum that is insensitive to standard gravity allows for this significant improvement over the current ISLV limit while using standard fabrication techniques.

One signature of ISLV is that the Laplacian of the Yukawa potential (\ref{potential}) is nonzero and could, in principle, be detected.  However, the spherical symmetry of the Laplacian renders it incapable of producing a torque on a torsion device.  A gradient of the Laplacian, such as $\tfrac{\partial}{\partial x}\nabla^2 V$, has a preferred direction and is capable of producing a torque.  Therefore, for this experiment, a nonzero horizontal gradient of the Laplacian is the ISLV signature of choice.

To gain insight into the nature of this putative interaction, consider the interaction energy of the pendulum and the source
\begin{equation}\label{energy}
U = \int \rho Vd^3r
\end{equation}
where $\rho$ is the density of the pendulum and $V$ is the potential generated by the source mass.  A Cartesian Taylor expansion of the potential about the origin (the center of mass of the pendulum) yields
\begin{equation}\label{energyexp}
U = \int \rho \left( 1 + x\frac{\partial}{\partial x} + y\frac{\partial}{\partial y} + z\frac{\partial}{\partial z} + \frac{1}{2} x^2 \frac{\partial^2}{\partial x^2} + xy \frac{\partial^2}{\partial x \partial y} + \ldots \right)V dxdydz.
\end{equation}
Collecting these terms into expressions with symmetries identical to the spherical harmonics\cite{berg} provides a convenient basis for a multipole representation given by
\begin{equation}
U = \sum_{n,l,m}M_{nlm}V_{nlm}
\end{equation}
where
\begin{equation}
M_{nlm}\propto \int \rho r^n Y_l^m d^3r
\end{equation}
and $V_{nlm}$ is the appropriate derivative of the potential.  The ``Newtonian'' terms, where $n=l$, satisfy Laplace's equation and can describe Newtonian gravity.  In contrast, ``non-Newtonian'' terms, where $n \neq l$, do not satisfy Laplace's equation (the non-Newtonian terms are typically ignored in the multipole expansions presented in textbooks for this reason).  The horizontal gradient of the Laplacian corresponds to the multipole term
\begin{equation}
\begin{split}
U_{311} &= V_{311}M_{311} \\
&= \frac{\partial}{\partial x} \left( \nabla^2 V \right) \int \rho x \left(x^2+y^2+z^2 \right) dxdydz \\
& \propto \frac{\partial}{\partial x} \left( \nabla^2 V \right) \int \rho r^3 Y_1^1 dxdydz
\end{split}
\end{equation}
where $M_{311}$ and $V_{311}$ are called the 311 mass and field moments respectively.  The 311 mass moment is manifestly non-Newtonian because the field moment to which it couples, $V_{311}$, is a derivative of the Laplacian and would be identically zero if the interaction were purely Newtonian.

In this experiment, the ratio of the size of the pendulum to the distance to the source mass is less than one.  This effectively supresses the contribution to the interaction energy from each higher multipole-order by that ratio.  The design, therefore, eliminates the relevant Newtonian mass and field moments from the pendulum and source mass up to the order where the contribution from the nonzero moments fall below estimated systematic errors.  In particular, since the signal has $m=1$ symmetry, all $m=1$ Newtonian mass and field moments from $l=1$ to $l=6$ are nulled by design.

Visualizing the shape of an object that is insensitive to Newtonian gravity and yet sensitive to deviations from Newtonian gravity is difficult.  Finding a place to begin the design of such an object is equally challenging.  Ultimately, the point of departure came from group theory; the $2l+1$ Newtonian moments for a given $l$ form a rotation group.  Consequently, the design of the pendulum began as five ideal ring masses placed along a common, vertically oriented symmetry axis.  The cylindrical symmetry of the ring masses ensures that only moments with $m=0$ symmetry are nonzero.  By properly placing the ring masses, the $m=0$ Newtonian moments for desired values of $n=l$ can be nulled.  Due to the properties of the rotation group, this eliminates \emph{all} Newtonian moments for that $n$.  The masses, vertical positions, and radii of these rings give 15 free parameters used to simultaneously maximize the non-Newtonian 310 mass moment, eliminate the leading Newtonian moments, and constrain the overall dimensions.  A rotation of the pendulum by $90^{\circ}$ about a horizontal axis transforms the 310 moment into a 311 moment, which can couple to the horizontal gradient of the Laplacian, while the low-order Newtonian moments remain zero.

\begin{figure}[!ht]
\begin{center}
\includegraphics[width=0.4\textwidth]{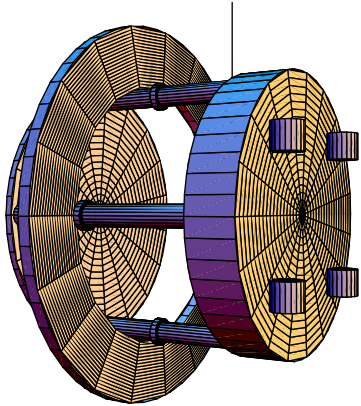}
\caption{Rendition of the ISLV pendulum.  The pendulum is made of fused silica parts with a total mass of 240g.  The diameter is roughly 8cm.  Four moveable trim masses are visible on surface of the thick disk.}
\label{pendulum}
\end{center}
\end{figure}

Once the basic mass distribution of the pendulum was determined, cylindrically symmetric connecting surfaces made the pendulum physically realizable.  Ultimately, reducing the machined surface area necessitated breaking the cylindrical symmetry in favor of a connecting structure with four-fold symmetry (see Figure \ref{pendulum}).  Appropriately adjusting available design parameters nulled the resulting $m=4$ moments.

\begin{figure}[!ht]
\begin{center}
\includegraphics[width=0.4\textwidth]{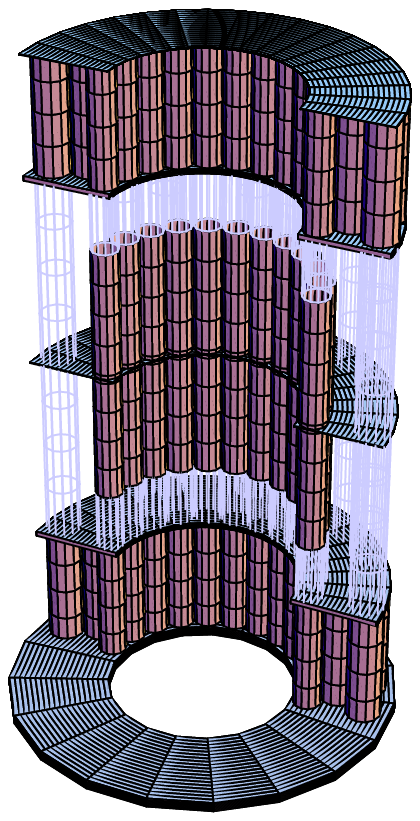}
\caption{Rendition of the ISLV source mass.  The mass is composed of several hundred precision machined solid stainless steel cylinders.  The wireframe components are hollow steel tubes.  The diameter of the base is approximately 1m.}
\label{source}
\end{center}
\end{figure}

The design of the ISLV source mass (Figure \ref{source}), similar to that the pendulum, eliminates all $m=1$ Newtonian derivatives of the potential below $l=8$.  Source mass top/bottom symmetry about the horizontal mid-plane naturally nulls the even $l$, $m=1$ potentials.  Manipulating the remaining degrees of freedom nulls the 331, 551, 771, and 220 potentials.  Eliminating the 220 potential renders the source mass insensitive to small tilts about the mid-plane that would otherwise generate an unwanted 221 potential.

Once constructed, the pendulum is suspended in the midplane of the source mass and set into large amplitude torsional oscillation.  The source mass rests upon an air bearing and is periodically lifted and rotated to one of several static positions in azimuth.  A signal that follows the source mass from position to position with an $m=1$ variation constitutes a detection of ISLV.  Such a signal could be the result of interactions with a scalar or vector partner to the graviton or an interaction via an unknown fifth force which couples approximately to mass\cite{paul,ephr}.

Historically, experiments designed to detect ISLV using a torsion device have relied on a pendulum with an exaggerated Newtonian moment.  Indeed, a barbell with a large $l = 2$, $m = 2$ mass moment achieves a high sensitivity to ISLV.  However, this exaggerated Newtonian moment couples to the residual $l = 2$, $m = 2$ field moments arising from fabrication errors associated with the source mass.  This interaction produces a large $m = 2$ systematic effect that would mimic a putative ISLV signal.  Thus, any scheme to increase the ISLV signal by exaggerating a Newtonian moment in the pendulum necessarily increases the limiting systematic effect as well.

Using a pendulum that has no low-order Newtonian moments reduces such systematic effects by two orders of magnitude.  Residual Newtonian moments resulting from fabrication errors in the source mass would not directly interact with the pendulum since it is specifically designed not to detect them.  It is therefore only the residual Newtonian mass moments resulting from errors in fabricating the pendulum that can couple to those residual Newtonian field moments.  Therefore, departures from the design configuration are manifest only in second order in these errors.  This design strategy dramatically reduces the systematic effects arising from standard machining tolerances, metrology, and material density inhomogeneity without resorting to heroic fabrication techniques\cite{berg}.

The experiment presented in this work is expected to constrain the strength of a putative inverse-square law violating ``fifth force'' to less than $\alpha = 10^{-5}$ at a length scale of $\lambda = 10^{-1}$m.  This constitutes a two order-of-magnitude reduction over the current limit at that scale.  Designing a pendulum that is virtually insensitive to Newtonian gravity and yet maximally sensitive to ISLV and a source mass that produces no low-order Newtonian gravitational fields and yet exaggerates the non-Newtonian horizontal gradient of the Laplacian renders the experiment only second-order sensitive to its leading systematic effect, fabrication errors in the pendulum and source mass.

\end{document}